\documentclass[usenatbib]{mn2e}
\usepackage{epsfig}

\def\I{{\sc i}}
\def\II{{\sc ii}}
\newcommand{\vsini}{\mbox{$v\sin\!i$}}

\begin{document}

\title[Abundance analysis of Am binaries and search for tidally driven 
abundance anomalies - III.]
{Abundance analysis of Am binaries and search for tidally driven   
abundance anomalies - III.
HD\,116657, HD\,138213, HD\,155375, HD\,159560, HD\,196544 and HD\,204188} 
\author[Stateva et al.]
{I. Stateva$^{1}$\thanks{E-mail: stateva@astro.bas.bg}, I.Kh. Iliev$^{1}$, J. Budaj$^{2}$ \\
$^1$Institute of Astronomy with NAO, Bulgarian Academy of Sciences, Sofia 1784, Bulgaria\\
$^2$Astronomical Institute, Slovak Academy of Sciences,
059 60 Tatransk\'{a} Lomnica, The Slovak Republic}

\date{Accepted~~~; Received~~~}

\maketitle
\begin{abstract}
We continue here the systematic abundance 
analysis of a sample of Am binaries in order to search for possible 
abundance anomalies driven by tidal interaction in these binary systems.

New CCD observations in two spectral regions (6400--6500, 6660--6760~\AA\AA) 
of HD\,116657, HD\,138213, HD\,155375, HD\,159560, HD\,196544 and HD\,204188
were obtained. Synthetic spectrum analysis was carried out and basic 
stellar properties, effective temperatures, gravities, projected rotational 
velocities, masses, ages and abundances of several elements were determined.
We conclude that all six stars are Am stars.

These stars were put into the context of other Am binaries with
$10 < P_{\rm orb} < 200$ days and their abundance anomalies discussed in 
the context of possible tidal effects. 
There is clear anti-correlation of the Am peculiarities with 
$v\sin i$.
However, there seems to be also a correlation with the
eccentricity and may be with the orbital period.
The dependence on the temperature, age, 
mass, and microturbulence was studied as well. The projected rotational 
velocities obtained by us were compared to those of 
\citet{rgbgz02} and \citet{am95}.
\end{abstract}


\section{Introduction}
The Am stars is a well known subgroup of chemically peculiar (CP) stars
on the upper Main Sequence (MS). They exhibit abnormally 
strong metallic and unusually weak Ca and Sc lines and, as a consequence,
the spectral types inferred from calcium lines are usually earlier than 
those from hydrogen lines, and the latter are earlier than the spectral 
types from the metallic lines. The anomalous 
intensity of most of these absorption lines is due to the abnormal 
chemical composition of superficial layers. 
The typical abundance pattern of Am stars is that
they exhibit a deficit of light elements like C, Mg, Ca, Sc
and progressively increasing overabundances of iron group and heavier 
elements. This abundance pattern is often referred to as the Am phenomenon.
Rotation was found to play a key role in these stars and
there is a growing amount of recent observational evidence  
that Am peculiarity is either a smooth or a step function of rotation 
(\citealt{ib08} and the references herein; \citealt{Abt00}; \citealt{bc00}).
Nevertheless, Am peculiarity does seem to depend on evolutionary status
or age as well. It may (1) develop very quickly soon after the star
arrives on the MS, or even before that 
(\citealt{bc00}), and does not undergo considerable changes 
during the MS phase, or (2) observable abundances of some elements may
vary with age and this can be used to constrain the evolutionary models
(\citealt{mr04}; \citealt{Monier05}). At the same time, no significant 
correlation of the abundance anomalies with $\vsini$ were found by 
\citet{mr04}, and \citet{Monier05}.
Apart from that the Am phenomenon is apparently restricted to a well-defined
region of the MS in the HR diagram which implies its dependence on
atmospheric parameters such as effective temperature and gravity 
(\citealt{kn98}; \citealt{Hui-Bon-Hoa00}). 

The Am peculiarity seems to depend on the orbital elements in a binary system
as well. \citet{Budaj96}, \citet{Budaj97}, \citet{ibzbz98} studied 
$\vsini~{\rm vs}~P_{\rm orb}$, $e~{\rm vs}~P_{\rm orb}$, 
$\delta m_{1}~{\rm vs}~P_{\rm orb}$, $f(m)~{\rm vs}~P_{\rm orb}$, 
$\vsini~{\rm vs}~P_{\rm p}$, $\delta m_{1}~{\rm vs}~P_{\rm p}$,
where $P_{\rm orb}$ is the orbital period, $\delta m_{1}$ is a metallicity parameter
which shows the difference in the dereddened $m_{1}$ index of $uvby\beta$ photometry
between an Am star and a normal star of the same index, $f(m)$ is a mass function, 
and $P_{p}$ is the ``instantaneous" orbital period at 
periastron. The authors concluded that there are a number of subtle effects which are 
difficult to understand within the current framework of the rotation and 
atmospheric parameters as the only agents determining the Am star peculiarity.

Since Am stars are often found in binaries (\citealt{ngccu98}; 
\citealt{dmcg00}) they provide a unique opportunity to study 
the influence of a companion on the stellar hydrodynamics.

This forced us to study the Am peculiarity and the 
orbital elements and $\vsini$, mass, age as well. In order to explore the possible 
dependence of Am peculiarity on the orbital elements of the binary system we started 
a systematic spectroscopic investigation of Am stars. A few tens of Am binaries from 
Budaj (1996) were chosen to be studied in details. The following criteria were applied 
in order to compile the star's list: targets with the declination $\delta> -10^\circ$ 
and brighter than the 7-th magnitude in V-filter. In order to cover a full range of 
eccentricities and avoid strong synchronization effects we have chosen only stars with 
orbital periods 10$^{d} < P_{\rm orb} < 200^{d}$. No constraints were put on the 
rotational velocity. 

In this connection the studies of individual Am binaries
are very important (e.g., \citealt{fossati07}; \citealt{zzmi08}; \citealt{zzmkisrk09}; 
\citealt{Boffin10}; \citealt{hgs10}; \citealt{mzz10}; \citealt{qtc10}). Obtaining new 
orbital elements of many Am binaries are also needed to prove the dependence 
between orbital elements and the Am peculiarity (\citealt{Debernardi02}; \citealt{cp07};
\citealt{zhao07}; \citealt{fw10}).

This is the last in a series of papers (\citet{bi03}, hereafter Paper~I 
and \citet{ibfsr06}, hereafter Paper~II) aiming at studying together the Am peculiarity 
in multi-dimension parameter space involving the orbital elements, $\vsini$, mass, and age.

\section{Observations and sample stars}

\begin{table}
\caption[]{Log of observations: Spectrum number, date [dd.mm.yyyy], 
HJD (2450000+) of the beginning of the exposure,
effective exposure time [min], spectral region, heliocentric
radial velocity of the primary and its error [km\,s$^{-1}$].}
\begin{center}
\begin{tabular}{cccrcrr}
\hline
N & Date  &  HJD & Exp. & Reg. & RV1 & $\Delta$RV1 \\
\hline
\multicolumn{7}{|c|}{HD\,116657}\\
\hline
1& 04.01.2001 & 1913.554 & 40 & Ca & -11.1 & 2.6\\
2& 04.01.2001 & 1913.594 & 40 & Li & -11.1 & 2.8\\
\hline
\multicolumn{7}{|c|}{HD\,138213}\\
1& 10.06.2001 & 2071.325 & 45 & Ca & -12.4 & 1.7 \\
2& 09.06.2001 & 2070.357 & 45 & Li & -11.2 & 3.1 \\
\hline
\multicolumn{7}{|c|}{HD\,155375}\\     
\hline
1& 10.06.2001 & 2071.392 & 90  & Ca & 24.7 & 2.8\\
2& 21.08.2001 & 2508.300 & 115 & Ca & 19.1 & 1.9\\
3& 26.08.2002 & 2513.251 & 90  & Ca & 25.6 & 1.8\\
4& 28.08.2002 & 2515.292 & 40  & Ca & 24.7 & 1.5\\   
5& 22.09.2007 & 4366.270 & 60  & Ca & 8.4  & 2.8\\         
6& 23.09.2007 & 4367.245 & 90  & Ca & 10.7 & 1.9\\     
7& 24.09.2007 & 4368.229 & 60  & Ca & 14.9 & 2.0\\     
8& 09.06.2001 & 2070.430 & 75  & Li & 24.6 & 5.3\\ 
\hline
\multicolumn{7}{|c|}{HD\,159560}\\   
\hline
1& 10.06.2001 & 2071.453 & 40 & Ca & -23.7 & 2.5\\
2& 23.08.2007 & 4336.395 & 25 & Ca & -13.8 & 2.2\\
3& 25.08.2007 & 4338.442 & 20 & Ca & -12.7 & 2.9\\
4& 11.06.2001 & 2072.384 & 45 & Li & -24.6 & 2.7\\
\hline
\multicolumn{7}{|c|}{HD\,196544}\\     
\hline
1& 10.06.2001 & 2069.546 & 60  & Ca & -29.9 & 2.5\\
2& 28.08.2001 & 2150.278 & 100 & Ca & -7.2  & 3.0\\ 
3& 02.09.2001 & 2155.332 & 30  & Ca &  12.6 & 4.7\\
4& 18.05.2002 & 2412.560 & 40  & Ca & -19.9 & 2.4\\ 
5& 10.08.2000 & 1767.324 & 115 & Li &  21.1 & 4.1\\  
\hline
\multicolumn{7}{|c|}{HD\,204188}\\
\hline
1& 23.07.2000 & 1749.437 & 180 & Ca & -18.5 & 2.5 \\
2& 22.07.2000 & 1748.454 & 260 & Li & -8.0  & 2.9 \\     
\hline
\end{tabular}
\end{center} 
\label{t1}
\end{table}  

Our spectroscopic observations were carried out with the 2-m RCC telescope of
the Bulgarian National Astronomical Observatory in the frame of our
scientific project on Am stars in binary systems. 
As in the previos papers of the series we observed each star in two 
spectral regions: 6400--6500~\AA\AA\ (Ca) and 6660--6760~\AA\AA\ (Li). For more 
details about the processing procedures see these papers.

\begin{figure}
\epsfig{file=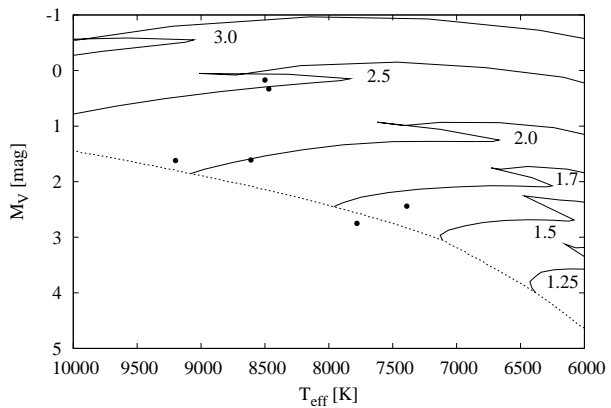,width=8.2cm,angle=0}
\caption{The location of our six stars in the HR diagram. 
Evolutionary tracks for 
$M=3.0, 2.5, 2.0, 1.7, 1.5,  ~{\rm and}~ 1.25~M_{\odot}$ 
are shown by solid lines, an isochrone for $\log T=3.0$ in [yr] (ZAMS)
is shown by the dotted line (\citealt{ls01}).}
\label{fhrd}
\end{figure}

The focus of this paper is to continue the analysis of another
six stars from the sample, namely HD\,116657, HD\,138213, HD\,155375, 
HD\,159560, HD\,196544 and HD\,204188.

\section{Atmospheric parameters and spectrum synthesis}

\begin{table*}
\caption{Photometry, atmospheric parameters and other relevant information
about the observed stars}
\begin{center}  
\begin{tabular}{|l|l|l|l|l|l|l|}
\hline
Star         &HD\,116657 &HD\,138213 &HD\,155375 &HD\,159560 &HD\,196544&HD\,204188 \\
\hline
\multicolumn{7}{|c|}{ UBV photometry}         \\
V            &2.227$^{T}$& 6.146 & 6.586 & 4.865 & 5.433 & 6.078 \\
B-V          &0.057$^{T}$& 0.10  & 0.085 & 0.279 & 0.052 & 0.22 \\
U-B          & --        & 0.12  & 0.086 & 0.068 & 0.039 & 0.06 \\
\hline
\multicolumn{7}{|c|}{$uvby\beta$ photometry}  \\
E(b-y)       &-0.010     &-0.014 &-0.005 & 0.001 &-0.010 & 0.000 \\
$(b-y)_{0}$  & 0.063     & 0.046 & 0.047 & 0.175 & 0.022 & 0.142 \\
$m_{0}$      & 0.239     & 0.191 & 0.197 & 0.208 & 0.186 & 0.199 \\
$c_{0}$      & 0.911     & 1.141 & 1.025 & 0.747 & 1.014 & 0.777 \\
$\beta$      & 2.886     & 2.860 & 2.885 & 2.772 & 2.911 & 2.806 \\
$T_{\rm eff}$& 8470      & 8430  & 8560  & 7460  & 9090  & 7780 \\
$\log g$     & 4.32      & 3.61  & 4.09  & 4.14  & 4.33  & 4.30 \\
\hline
\multicolumn{7}{|c|}{Geneva photometry}       \\
U            & --        & 1.648 & 1.561 & 1.478 & 1.501 & 1.434 \\
V            & --        & 0.825 & 0.842 & 0.609 & 0.905 & 0.677 \\
B1           & --        & 0.909 & 0.910 & 0.971 & 0.902 & 0.953 \\
B2           & --        & 1.445 & 1.449 & 1.404 & 1.469 & 1.422 \\
V1           & --        & 1.524 & 1.539 & 1.334 & 1.601 & 1.390 \\
G            & --        & 1.989 & 2.000 & 1.732 & 2.084 & 1.807 \\
$T_{\rm eff}$& --        & 8403  & 8657  & 7321  & 9301  & - \\
$\log g$     & --        & 3.65  & 4.02  & 4.26  & 4.29  & - \\
\hline
$P_{\rm orb}$& 175.6$^{1}$ & 105.95$^{2}$ & 23.2$^{3}$ & 38.0$^{4}$ & 11.0$^{5}$ & 21.72$^{5}$ \\
e            & 0.46$^{1}$  & 0.0$^{2}$    & 0.42$^{3}$ & 0.03$^{4}$ & 0.23$^{5}$ & 0.0$^{6}$ \\
\hline
$\pi$        &$41.73\pm0.61$ &$6.37\pm0.29$ &$10.13\pm0.45$ &$32.80\pm0.18$ &$17.26\pm0.33$ &$21.57\pm0.56$\\
$M_{V}$      & 0.33          & 0.17         & 1.61         & 2.44          & 1.62          & 2.75 \\        
\hline
\multicolumn{7}{|c|}{adopted atmospheric parameters}\\
\hline
$T_{\rm eff}$ & 8470 & 8500 & 8610 & 7390 & 9200 & 7780 \\  
$\log g$      & 4.32 & 3.50 & 4.06 & 4.20 & 4.31 & 4.30 \\
\hline
\end{tabular}
\end{center} 
Note: 
$^{1}$ -- \citet{Gutmann65}; 
$^{2}$ -- \citet{ls71};
$^{3}$ -- \citet{Debernardi02}; 
$^{4}$ -- \citet{margoni92};
$^{5}$ -- \citet{Harper35};
$^{6}$ -- \citet{bfm78};
$^{T}$ -- Hipparcos and Tycho catalogue (\citealt{ESA97});
$T_{\rm eff}$ is in [K], $\log g$ in CGS units, $P_{\rm orb}$ in days and 
$\pi$ in [mas].
\label{t2}   
\end{table*} 

Relevant information about our program stars is summarized in Table \ref{t2}.
The $uvby\beta$ indices (de-reddened using the UVBYBETA code of 
\citealt{md85}) were taken from \citet{Renson91}.
Geneva and UBV photometry were from \citet*{mmh97}. 
The improved Hipparcos parallaxes were taken 
from \citet{vanLeeuwen07}.
Table \ref{t2} also lists the absolute $M_{V}$ magnitudes obtained from these 
parallaxes and V photometry.
The atmospheric parameters were derived from both $uvby\beta$ and Geneva 
photometry. If both estimates were available we accepted
their rounded mean as the best choice for model atmosphere parameters.
All these stars seem to be SB1 binaries or have only a very weak
secondary spectrum, hence the possible influence of their companions on 
photometry was neglected.

A detailed spectrum synthesis of the spectral regions was accomplished 
following the same recipe as in the previous papers of the series.

\begin{figure*} 
\epsfig{file=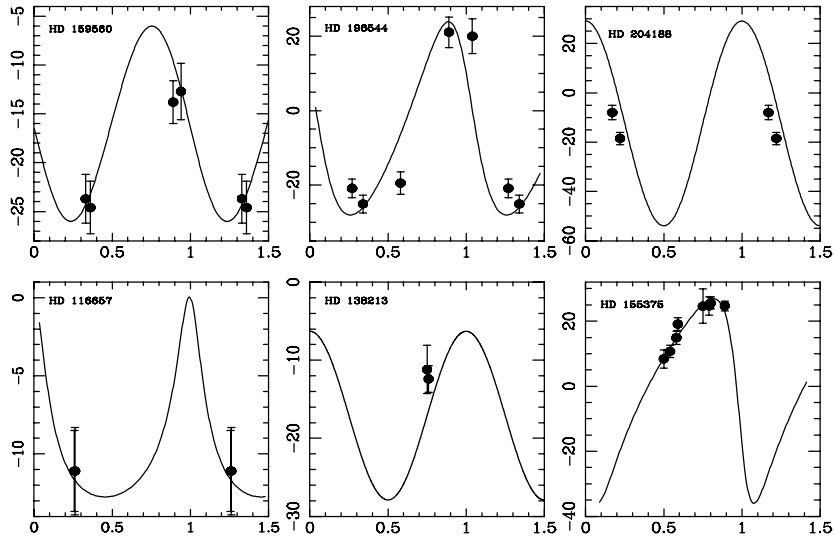, width=11cm}
\caption{Measured radial velocities as listed in Table \ref{t1} (dots) 
in comparison with the predicted radial velocity curves (solid lines) 
versus phase. Orbital elements are taken from the appropriate references 
shown in Table \ref{t2}.}
\label{fnew}
\end{figure*}

\begin{figure*}
\epsfig{file=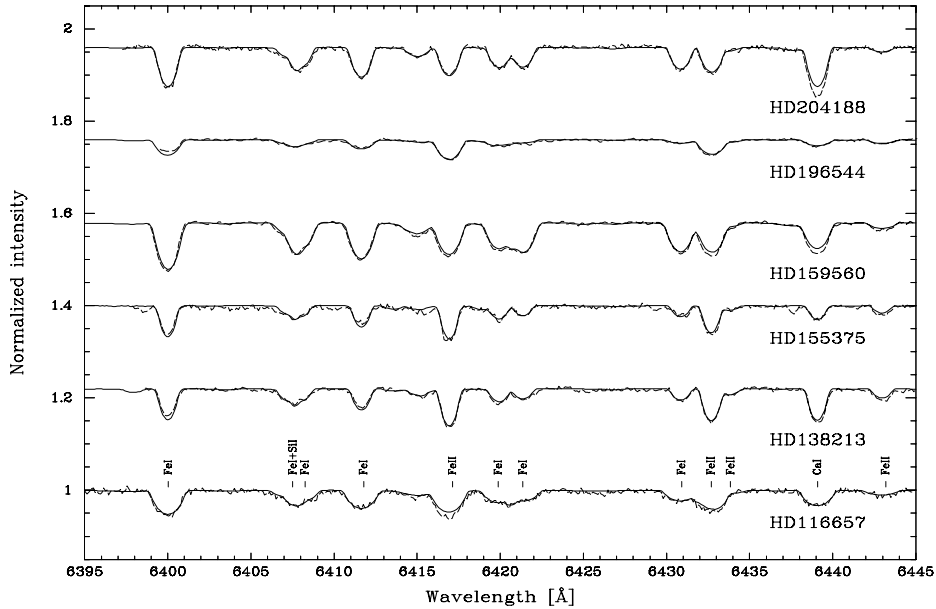,width=13cm,angle=0}
\caption{Synthetic (solid) and observed (dashed) spectra of all of our 
programme stars. The observed spectra were shifted to the laboratory
frame of synthetic spectra. A proper shift to the vertical coordinate
was applied to all but the lowest spectrum.}
\label{fsyn1}  
\end{figure*}

\section{Results for individual stars}
The abundances obtained by synthetic spectrum fitting analysis are 
expressed relative to the Sun in terms of 
${\rm [N/H]=\log (N/H)_{\star}-\log (N/H)_{\odot}}$
in the Table \ref{t5}.
Taking into account the accuracy of the atmospheric parameters, as
well as the atomic data, the abundances of Al, Si, S, Ca and Fe are 
generally determined within $\la 0.2$\,dex, while the abundances
of the other elements, which mainly occur in weak blends, 
are only approximate. The Ba abundances should be used with caution
as they are usually derived from only one line (Ba\,\II\, 6497~\AA),
which is at the edge of our frames. Apart from abundances and atmospheric 
parameters, we also derived the basic stellar properties like mass and age. 
We used the determined absolute magnitudes, created an HR diagram, and 
interpolated the evolutionary tracks and isochrones of \citet{ls01}. The 
masses, ages, and expected terminal-age main sequence (TAMS) obtained are 
also listed in Table \ref{t5} and the position of our programme stars in 
the HR diagram is illustrated in Figure \ref{fhrd}. Both the synthetic and 
observed spectra are depicted in Figure \ref{fsyn1}. 

Radial velocities, projected rotational velocities and microturbulent 
velocities determined as by-products are listed also in the 
Tables \ref{t1} and \ref{t5}. A comparison between measured radial 
velocities of the six stars and predicted radial velocity curves is shown 
in Figure \ref{fnew}. For the Am stars with $\vsini<50$ km\,s$^{-1}$
the radial velocities of the primary stars were measured using 
the cross-correlation of the whole spectral region of the observed spectrum 
with the synthetic spectra. For the fast rotating star 
(HD\,116657) with $\vsini>50$ km\,s$^{-1}$ the cross-correlation technique
produces large errors due to heavy line blending in some spectral regions.
Consequently, in the Li-region, we restricted the cross-correlation
region to 6710--6765~\AA\AA, and in the Ca region, we measured the velocities 
from the Ca\,\I\, 6439~\AA\ line using the center of mass method.
The measurements of \citet{bi03} with the same telescope configuration 
and center of mass method demonstrated, on the example of the fast rotating 
star (HD\,178449, $\vsini=139~$km\,s$^{-1}$), that the standard deviation
of such radial velocity measurements was less than 2 km\,s$^{-1}$. 
A discussion of the individual stars follows.

\begin{table*}
\caption{Abundances derived in terms of [N/H] for our six stars.
Abundances of the Sun are in terms of $\log (N_{\rm el}/N_{\rm H}) + 12.00$}
\begin{center}
\begin{tabular}{|l|l|l|l|l|l|l|l|}
\hline
&\multicolumn{1}{c|}{Sun}&\multicolumn{1}{c|}{HD\,116657}
&\multicolumn{1}{c|}{HD\,138213}&\multicolumn{1}{c|}{HD\,155375}
&\multicolumn{1}{c|}{HD\,159560}&\multicolumn{1}{c|}{HD\,196544}
&\multicolumn{1}{c|}{HD\,204188}\\
\hline
Li & 1.10 &$\leq +2.08$ &$\leq +2.2$  &$\leq +1.88$ &      +1.68  &$\leq +2.4$  &$\leq +0.6$ \\
C  & 8.52 &$\leq -0.62$ &$\leq -0.18$ &$\leq -0.11$ &      -0.57  &$\leq -0.34$ & -0.82 \\
O  & 8.83 &$\leq -0.38$ & -0.13       &$\leq -0.22$ &$\leq -0.16$ &$\leq -0.29$ & -0.13 \\ 
Al & 6.47 &      +0.30  & --          &       --    &      --     &     --      & -- \\
Si & 7.55 &      +0.00  & +0.34       &      +0.05  &      +0.01  &     +0.10   & +0.04 \\
S  & 7.33 &      +0.10  & +0.19       &      +0.0   &      +0.03  &     +0.15   & -0.31 \\
Ca & 6.36 &      -0.42  & +0.11       &      -0.64  &      -0.76  &     -0.50   & -0.18 \\
Ti & 5.02 &      +0.00  & +0.28       &      +0.14  &      -0.04  &     +0.24   & +0.21 \\
Fe & 7.50 &      +0.34  & +0.29       &      +0.22  &      +0.29  &     +0.27   & +0.07 \\
Ni & 6.25 &      +0.67  & +0.40       &      +0.65  &      +0.51  &     --      & +0.23 \\
Ba & 2.21 &      --     & +2.15       &      +1.85  &      +1.64  &     +1.19   & +1.41 \\
\hline
$\xi_{\rm turb}$  &-- & 2.0 & 2.0 & 2.1 & 2.7 & 2.4 & 2.0 \\
$\vsini$          &-- & 51  & 32  & 31  & 42  & 43  & 36  \\
\hline
$M$               &-- & 2.10 & 2.49 & 2.26 & 1.62 & 2.20 & 1.67 \\
$\log T$          &-- & 8.72 & 8.43 & 8.05 & 8.96 & 7.43 & 6.5$^{*}$ \\
$\log TAMS$       &-- & 8.77 & 8.8 & 8.99 & 9.32 & 9.02 & 9.28 \\ 
\hline
\end{tabular}
\end{center} 
Note: 
Sun -- abundances are taken from \citet{gs98}
(recall that normal lithium abundance in hot stars or
meteorites is [Li/H]=2.00);
microturbulence - $\xi_{\rm turb}$ and $\vsini$ are in km\,s$^{-1}$,
M is mass in $M_{\odot}$,
the age T and the Terminal-Age-Main-Sequence TAMS are given in years; $^{*}$ - the age is only upper 
limit (see details in the text)
\label{t5}
\end{table*}

\begin{figure*}
\epsfig{file=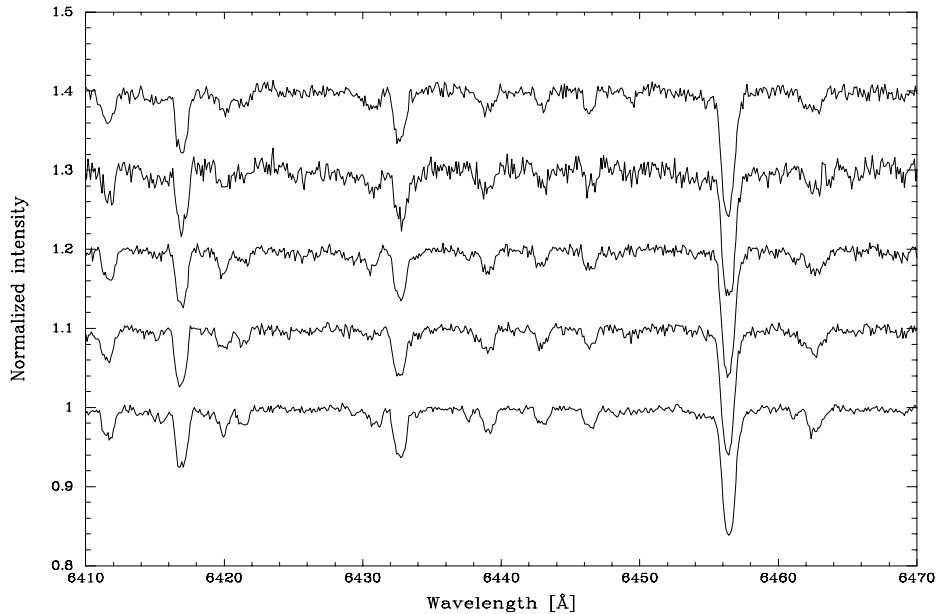,width=13cm,angle=0}
\caption{Observed spectra of HD\,155375. A proper shift to the vertical coordinate
was applied to all but the lowest spectrum. Spectra with low S/N are not shown in the figure.}
\label{fhd155}  
\end{figure*} 

\subsection{HD\,116657}
HD\,116657 ($\zeta$ UMa B, HR\,5055, ADS\,8891 B, BD\,+55 1598B, 
SAO\,28738, CSV\,101382, A1m) is a part of the Mizar system. This system 
has been the first double star observed photographically. HD\,116657 is
the fainter star of the system where both stars are spectroscopic binaries.

The first spectral classification was given by \citet{Roman49} - A2/A8/A7 
(Ca\,II/H/metal lines). Later \citet{ccjj69} determined the spectral class as A1  
from K-lines and \citet{la78} as A2/A7/A9 from K/H/metallic lines. Finally 
\citet{al85} determined the spectral class of HD\,116657 as A1/A4/A3 according to 
K/H/metallic lines. The orbital elements of the binary system were obtained by 
\citet{Gutmann65} - $P_{\rm orb}=175.6^{d}$, $K=6.4~{\rm km\,s^{-1}}$,
\textit{e}=0.463, $V_0=-9.3~{\rm km\,s^{-1}}$, $\omega={\rm 6^{\circ}.9}$. 
There have been numerous evaluations of the projected rotational velocity given by
different authors. \citet{meadows61} and \citet{am73} measured 
$\vsini=50~$km\,s$^{-1}$. \citet{dobrichev85} gave close value
- $\vsini=52~$km\,s$^{-1}$ while \citet{Slettebak54} obtained higher velocity - 
$\vsini=75~$km\,s$^{-1}$. \citet{am95} measured the projected rotational 
velocity as $\vsini=51~$km\,s$^{-1}$ but scaling this value to the results of 
\citet{rgbgz02} changed the velocity to $\vsini=61~$km\,s$^{-1}$. \citet{Monier05} 
obtained also $\vsini=61~$km\,s$^{-1}$. Our value of $\vsini=51~$km\,s$^{-1}$ is 
very close to the values given by the majority of the authors. 
A few evaluations of the effective temperatures have been noted in the literature.
The first value given by \citet{Strobel60} determined HD\,116657 as a very cool 
star - $T_{\rm eff}$ = 5130\,K. The later values have given higher temperatures for the
star. According to \citet{Smith71} the effective temperature was $T_{\rm eff}$ = 8800\,K
and according to \citet{king03} and \citet{Monier05} - $T_{\rm eff}$=8425\,K. Our value
of $T_{\rm eff}$=8470\,K is very close to these authors' value.

The abundances of Ca, Fe, S are determined with good precision. 
For lithium we can only set an upper limit and claim that Li is not 
overabundant relative to the cosmic Li abundance.
However, there is an indication of Li line. If this is confirmed then 
\citet{ibzbz98} Li abundance would be underestimated. The values listed
for C and O are only upper limits also. According to our analysis Al is 
overabundant and Si seems normal from only one line SiI $\lambda$6721\,\AA\,. Ti
could be evaluated only approximately to be normal from two blends of CaI $\lambda$6717\,\AA\, 
and FeI $\lambda$6678\,\AA. Ni is found overabundant.
Some FeII lines are stronger than expected and it might indicate not only 
higher microturbulence or effective temperature but also lower gravity.

\subsection{HD\,138213}
HD\,138213(HR\,5752, HIP\,75770, BD\,+ 47 2227) is a spectroscopic binary star. It was classified
for the first time as a marginal metallic line star A5m: by \citet{ccjj69}. Later \citet{eggen76} 
determined HD\,138213 as a possible Am star based on photometry and \citet{floquet75} suggested that 
the star was spectroscopically variable. According to \citet{am95} the star was A2\,IV. The orbital 
elements were taken by \citet{ls71} - $P_{\rm orb}=105.95^{d}$, $K=10.8~{\rm km\,s^{-1}}$,
\textit{e}=0, $V_0=-17.10~{\rm km\,s^{-1}}$, $\omega={\rm 0^{\circ}}$. Our radial velocities 
do not agree very well with the predicted velocity curve. It might be due to a small phase shift which 
accumulated over the years. 
There have been a few evaluations of the projected rotational velocity. \citet{abt75} gave 
$\vsini=30~$km\,s$^{-1}$ but later corrected
it to $\vsini=45~$km\,s$^{-1}$ (\citet{am95}). \citet{rgbgz02} scaled this value to their system and 
determined the velocity as $\vsini=54~$km\,s$^{-1}$. Our value is closer to the result given by \citet{abt75} 
- $\vsini=32~$km\,s$^{-1}$.

Our analysis confirms that HD\,138213 is a marginal Am star - Fe is overabundant and 
O is underabundant. Ca is almost solar abundant. For C and Li we give only the upper limits. 

\subsection{HD\,155375}
HD\,155375(HR\,6385, BD\,+12 3161, HIP\,84036, SAO\,102632, A1m) is a
spectroscopic binary star. \citet{osawa58} determined the spectral class of the star as
A1/A3/A5 from K/H/metallic lines and \citet{ccjj69} - as A1. Later \citet{am95} specified
it as A2\,III class and \citet{pdhkw01} - as A3\,V class. \citet{am95} also determined the projected
rotational velocity as $\vsini=25~$km\,s$^{-1}$ but scaling this value to the system of 
\citet{rgbgz02} the latter changed the velocity to $\vsini=33~$km\,s$^{-1}$. Our value of 
$\vsini=31~$km\,s$^{-1}$ is in good agreement with \citet{rgbgz02} result.
We used the orbital elements given by \citet{Debernardi02}: $P_{\rm orb}=23.25^{d}$, 
$K=31.42~{\rm km\,s^{-1}}$, \textit{e}=0.422, $V_0=1.03~{\rm km\,s^{-1}}$, $\omega={\rm 114^{\circ}.86}$. 
The radial velocities measured from our spectra are in excelent agreement with the radial 
velocities calculated by using these elements (see Figure \ref{fnew}).

There are many Fe lines in the spectral region of 6400-6500\,\AA\AA\,, so the abundance of Fe is very well
determined. Ca is underabundant. The abundances given at Table \ref{t5} for C and O are upper limits.
The situation with Li is the same as in the case of HD\,116657 - the Li line is very weak and the obtained Li
abundance is only upper limit. 

More than one spectrum have been obtained in order to check the possible variability of some lines in the 
spectrum of HD\,155375. As it is seen, a few lines have changed their profiles (see Figure \ref{fhd155}).
The relative changes of two lines, FeI $\lambda$6419.95\,\AA\, and FeI $\lambda$6421.35\,\AA, have been most 
obvious. Two calcium lines, CaI $\lambda$6439.08\,\AA\, and CaI $\lambda$6462.57\,\AA, have shown changes, too. 
The center of the lines has been changed and also there could be seen some features emerging from the blue 
side of the lines. All these observable clues forced us to suspect HD\,155375 as a new SB2 star. 

\subsection{HD\,159560}
HD\,159560 ($\nu^2$ Dra, HR\,6555, BD\,+55 1945, HIP\,85829, SAO\,30450, ADS\,10628 A, A4m) 
is a member of the visual binary system. The angular separation between the components is ${\rm 61.9}\arcsec$. 
Both components of the binary system are Am stars.
There have been many determinations of the stellar spectral class in the literature. The first evaluation 
was given by \citet{Slettebak49} - A2/F0/F5\,IV from K/H/metallic lines. Later the author specified the 
spectral class from H-lines as A7 (\citealt{Slettebak63}). According to \citet{ac84} the star was A4/F2V/F3.
\citet{ccjj69} also determined the spectral class of HD\,159560 as A4 from H-lines. 
The evaluations of the projected rotational velocity of the star have been very different. 
\citet{am73} gave $\vsini=35~$km\,s$^{-1}$, \citet{bvd80} - $\vsini=47~$km\,s$^{-1}$, \citet{al85} - 
$\vsini=50~$km\,s$^{-1}$, \citet{am95} - $\vsini=58~$km\,s$^{-1}$. Finally, \citet{rgbgz02} determined the 
rotational velocity of the star as $\vsini=68~$km\,s$^{-1}$ scaled from the result of \citet{am95}. We
obtained $\vsini=42~$km\,s$^{-1}$ and this is the value of the rotational velocity we used for spectrum
synthesis.
We used the two different set of orbital elements given by \citet{al85} and \citet{margoni92} in order 
to check our radial velocities. Our results are in good agreement with the radial velocity curve obtained by 
using the orbital elements of the latter authors (see Figure \ref{fnew}). They gave the following orbital elements 
- $P_{\rm orb}=38.034^{d}$, $K=10.0~{\rm km\,s^{-1}}$, \textit{e}=0.03, $V_0=-16.0~{\rm km\,s^{-1}}$, $\omega={\rm 92^{\circ}}$.

Our results of the abundances define the star as Am star. The line of Li\,I $\lambda$\,6707\,\AA\, is well 
identified and is relatively stronger comparing to the other stars in this investigation. That is why the 
obtained abundance of Li is well determined.
 
\subsection{HD\,196544}
HD\,196544 ($\iota$ Del, HR\,7883, BD\,+10 4339, HIP\,101800, SAO\,106322) 
is an A2V spectroscopic binary star. 
\citet{adams12} determined for the first time the stellar 
radial velocity and obtained that it was variable. He classified the star as A2 spectral 
type star. Later \citet{osawa59} determined the spectral class as A2\,V according to MK system and 
as A4 from metallic lines. \citet{Levato75} confirmed the spectral class as A2\,V and obtained 
the projected rotational velocity $\vsini=55~$km\,s$^{-1}$ but later revised the velocity to 
$\vsini=60~$km\,s$^{-1}$ (\citealt{gl84}). On the other hand, \citet{am95}
classified the star as A1\,IV and gave a smaller value of the rotational velocity -
$\vsini=30~$km\,s$^{-1}$. \citet{rgbgz02} derived the projected rotational velocity and after 
merging it with the data available obtained $\vsini=41~$km\,s$^{-1}$. The orbital elements were
derived by \citet{Harper35} - $P_{\rm orb}=11.039^{d}$, $K=26.0~{\rm km\,s^{-1}}$,
{\textit{e}=0.23, $V_0=-4.9~{\rm km\,s^{-1}}$, $\omega={\rm 61^{\circ}.8}$. Again, as in the case 
of HD\,138213 the differences between our radial velocity measurements and the predicted velocity
curve are due to a small phase shift accumulated over the years.

There have been a few elements abundances published in the literature. \citet{lemke89}, \citet{lemke90} 
and \citet{rh97} gave the abundances of Fe, Ti, C, Ba, N and S using the atmospheric parameters  
$T_{\rm eff}$=9100\,K and $\log g$=4.3 which were very close to our values (see Table \ref{t2}). 

According to the abundances obtained by us, HD\,196544 seems to be an Am star. The abundances of Fe 
and Ca are very well determined as Ca is underabundant and Fe is overabundant. Li line of 
6707\,\AA\, is very weak and the abundance of Li is determined as only an upper limit. Other elements like
C, O and Ti have weak lines in this spectral region, so their abundances should be scrutinized
as upper limits. The differences between the abundances of Fe, Ti and C obtained by us and published 
by \citet{lemke89} are within the errors.

\subsection{HD\,204188}
HD\,204188(IK Peg, HR\,8210, HIP\,105860, BD\,+18 4794, WD\,2124+191, A8m) is an interesting
single-lined spectroscopic binary with a companion star which is a massive white dwarf.
\citet{ccjj69} identified HR\,8210 as a marginal Am star and determined the spectral class as
A8m: but \citet{abtbid69} identified it as a definite Am star. According to \citet{bertaud70} the spectral
class of HR\,8210 was between A5 and F0. Later \citet{Guthrie87} in his study of the calcium abundances 
in metallic-line stars found that Ca was almost solar abundant. The most completed spectral identification
was made by \citet{am95} - A6/A9/F0 from K/H/metallic lines.
\citet{kurtz78} obtained that the primary Am star was also $\delta$ Sct star. Till now there are only a few 
stars which combine in one and the same star such contradictory characteristics.
The first orbit determination was made by \citet{Harper27} who found the period of about 27 days and
almost circular orbit. Later he clarified the period (\citet{Harper35}) and \citet{bfm78} assumed
the eccentricity as \textit{e}=0. We used the orbital parameters from the Ninth Catalogue of Spectroscopic Binary Orbits (SB9)(\citet{Pourbaix04}: $P_{\rm orb}=21.724^{d}$, 
$K=41.5~{\rm km\,s^{-1}}$, {\textit{e}=0, $V_0=-12.4~{\rm km\,s^{-1}}$, $\omega={\rm 0^{\circ}}$. 
There have been many evaluations of the projected rotational velocity of IK Peg in the literature -
from $\vsini=80~$km\,s$^{-1}$ (\citet{Levato75}) to $\vsini=31~$km\,s$^{-1}$ (\citet{am95}). \citet{rgbgz02}
determined the projected rotational velocity as $\vsini=40~$km\,s$^{-1}$ scaling the results of \citet{am95}.
The projected rotational velocity obtained by us $\vsini=36~$km\,s$^{-1}$ is in the range of \citet{am95} 
and \citet{rgbgz02} and very close to the values given by \citet{rll00}.

\begin{figure}
\epsfig{file=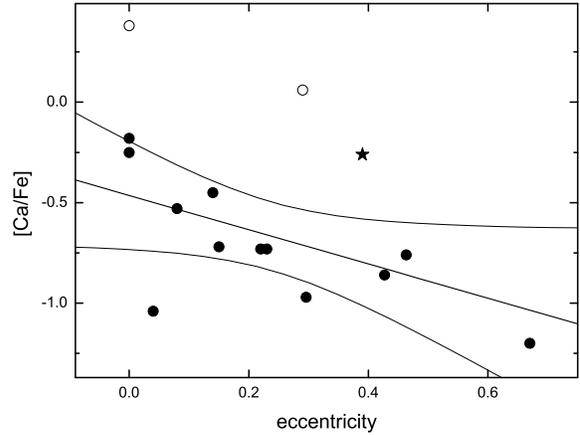, width=11.5cm}
\caption{[Ca/Fe] versus the eccentricity of the orbit. A least-squares regression line and 
the 95 percent confidence limits are drawn. The correlation coefficient is -0.55$\pm$0.05. HD\,198391 is 
denoted by an asterisk. Two non-Am stars HD\,178449 and HD\,18778 denoted by open circles are given only 
for completeness and not included in the analysis.} 
\label{feccentr}
\end{figure}

\begin{figure}
\epsfig{file=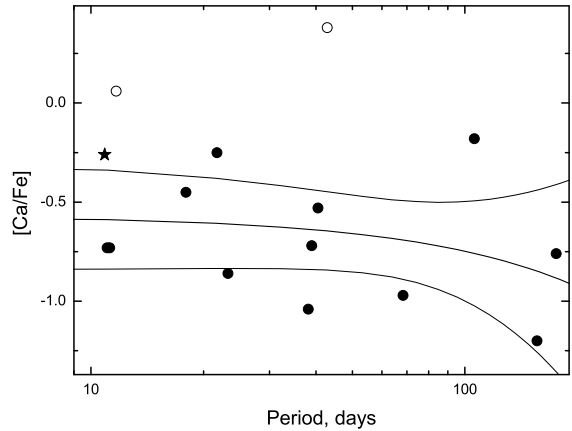, width=10.5cm}
\caption{[Ca/Fe] versus the orbital period. The periods are presented in logarithmic 
scale for more clarity. The correlation coefficient is -0.33$\pm0.06$.} The symbols are the 
same as in the previous figure.
\label{fperiod}
\end{figure}

The iron-peak elements like Fe, Ti and Ni are sligthly overabundant and those of Ca and O - underabundant. For Li we 
present only an upper limit. As for the other stars of this study Ba is overabundant. These results as well as the
atmospheric parameters are in good agreement with the results of \citet{sswa96}. 
Our abundance analysis confirms that HD\,204188 is a mild metallic line star.

The age obtained for this star is determined with not very high accuracy. We agree with \citet{kn98}
who noticed the difficulty of determing the stellar age properly when the star is near ZAMS.
 
\section{Tidal, rotation and evolution effects in Am phenomena}
Following the systematic search for abundance anomalies of Am stars driven by tidal effects we 
continued with studying the dependences between the chemical abundances and the orbital elements 
of the binary systems, the projected rotational velocity and the physical parameters of Am stars 
like mass, age, temperature. The main advantage of our analysis is the homogeneity of 
the observational material we used - all data were obtained at one telescope with the same spectrograph 
and detector as well as processed with the same data reduction package and analysed with the same
code Synspec.

As it is well known Am peculiarities are mainly manifested by Ca deficit and Fe overabundances 
and they could be represented in a reliable way through the values of [Ca/Fe] where 
[Ca/Fe]=[Ca/H]-[Fe/H]. This ratio would multiply the effect of Am peculiarities because these two 
elements have shown an opposite behaviour. Besides the lines of both elements get weaker with the 
increasing temperature and consequently [Ca/Fe] ratio is not that sensitive to the uncertainties 
of the effective temperature.
   
Up to now fifteen stars from our sample were fully processed. The results are taken from this work 
and also from Paper~I and Paper~II. Two of the stars in this sample, HD\,178449 and HD\,18778 
were found not to be Am stars. They are put on the figures but not included in the analysis.

First we investigate the dependences of [Ca/Fe] from the eccentricity (see Figure \ref{feccentr}). 
Despite of some scatter in the data there seems to be a trend like that mentioned earlier
by \citet{Budaj97} and \citet{ibzbz98}. The correlation coefficient is -0.55$\pm$0.05.
One star, HD\,198391 denoted by an asterisk in the figures, is distinguished from the common trend. 
This star is the hottest star amongst the sample and it is not a typical Am star (Paper~I). The star 
could be a transitional object between Am and the hotter HgMn stars because of its position in HRD 
close to the region occupied by HgMn stars and also of its abundances (for more details see 
Paper~I). Namely, that Am peculiarities increase ([Ca/Fe] decreases) with increasing eccentricity 
of the binary system. The dependence of [Ca/Fe] from the orbital period 
shown in Figure \ref{fperiod} is not so clear but still there is a tendency the metalicity to increase
towards the longer periods. Of course, these two parameters characterizing the binary system, eccentricity
and period, are not fully independent because of the synchronization and circularization of the orbits. 
This was the main reason to analyse only stars with periods between 10 and 200 days. The comparison of  
the three parameters simultaneously in 3D-graph (see Figure \ref{feperiod}) - metalicity, eccentricity and 
period showed us that there was not any correlation between period and eccentricity in the whole range of periods 
which extends even beyond 150 days.

\begin{figure} 
\epsfig{ file=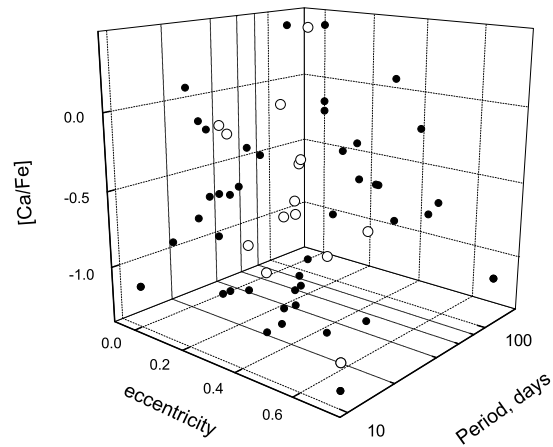,origin=tr,width=9.2cm}
\caption{The dependence between [Ca/Fe], the eccentricity and the orbital period. Open circles 
stand for plane projections.}
\label{feperiod}
\end{figure}

We study also the dependence of the Am peculiarities on the projected rotational velocity of the Am stars 
(see Figure \ref{fvsini}). In general, there is a clear trend of increasing the peculiarity 
(decreasing of [Ca/Fe]) towards small values of $\vsini$ - the correlation coefficient is 
+0.85$\pm0.04$. Besides the hottest star already mentioned, HD\,198391, two other stars, HD\,138213 and 
HD\,204188, both marginal Am stars, do not follow the common trend. The fact that these two stars 
depart from the clear smooth correlation of metallicity and $\vsini$ indicates that the rotation 
is not the only agent responsible for this peculiarity. It confirms our claims from \citet{ibzbz98}
that the low Am peculiarity in these two systems is due to their small eccentricity. Again in order 
to check the possible correlation between the parameters studied we tried to combine [Ca/Fe], the
eccentricity and $\vsini$ in 3D-graph (see Figure \ref{fevsini}). We obtained a weak negative correlation 
between the eccentricity and the projected rotational velocity - faster rotating stars tended to have smaller
eccentricities. 

\begin{figure} 
\epsfig{ file=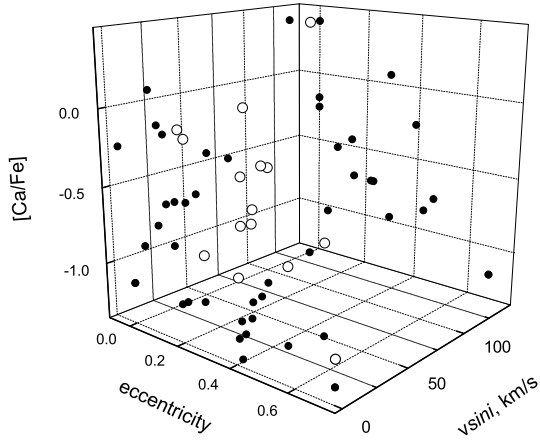,origin=tr,width=9.2cm}
\caption{The dependence between [Ca/Fe], the eccentricity and the projected rotational velocity. Again, 
open circles stand for plane projections.}
\label{fevsini}
\end{figure}

\begin{figure}
\epsfig{file=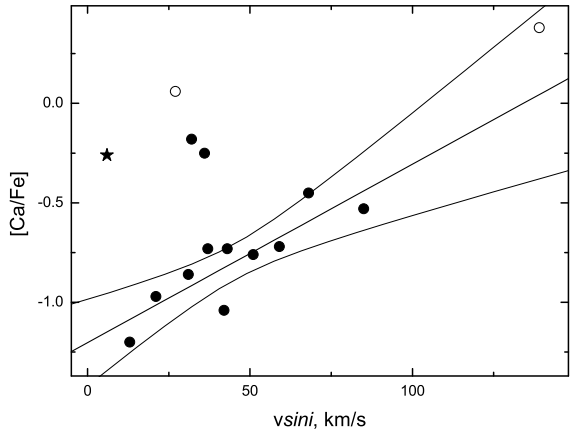, width=16.2cm}
\caption{[Ca/Fe] versus the projected rotational velocity. The correlation coefficient is +0.85$\pm0.04$.
The symbols are the same as in Figure \ref{feccentr}.} 
\label{fvsini}
\end{figure}

The dependence of the Am peculiarity on the effective
temperature is plotted on the Fig. \ref{fteff}. There might be a trend of 
decreasing the peculiarity with the temperature - the correlation coefficient is +0.36$\pm0.06$.

We sought a possible dependence of the Am peculiarity on the age and the mass of the 
stars. With the exception of one star, HD\,196544, the other stars studied have very 
close ages - the difference between the youngest and the oldest star is 0.51 in $\log T$ (where $T$ is the 
age in years).
The majority of the masses of the stars is also in very short interval - 0.5 in solar masses.
As a result from all our data we can not see any signs of dependence between these parameters 
(mass and age) and the Am peculiarity.

\begin{figure} 
\epsfig{ file=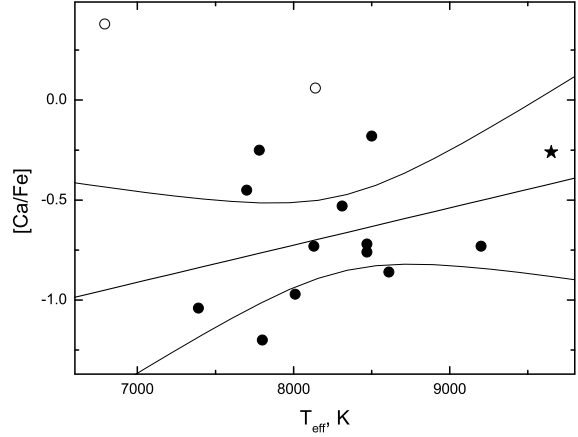,width=16.5cm}
\caption{[Ca/Fe] versus the effective temperature. The correlation coefficient is +0.36$\pm0.06$.
The symbols are the same as in Figure \ref{feccentr}.}
\label{fteff}
\end{figure}

We also studied the connection between the microturbulence and the effective temperature 
(see Fig. \ref{fvturb}). 
There seems not to be any dependence between these parameters for the temperature 
region 7000-9000\,K. But at higher temperatures it is possible to claim that the microturbulence 
decreases with the increasing temperature. These results fit relatively well with the conclusion 
given by \citet{bc92}. 
The microturbulence is a pure fitting parameter which brings into 
agreement the abundances from weak and strong lines. It does not necessarily have 
the meaning of existing turbulent motions. Nevertheless, this behaviour is in
agreement with the expectations and with the fact that supeficial convective
zones get thinner at higher temperatures and cease at about 10\,000\,K.
We also studied a possible dependence of the Am peculiarities on the
microturbulence but there is no clear correlation. Apparently, 
the microturbulence does not seem to destroy the Am peculiarity.

\begin{figure} 
\epsfig{ file=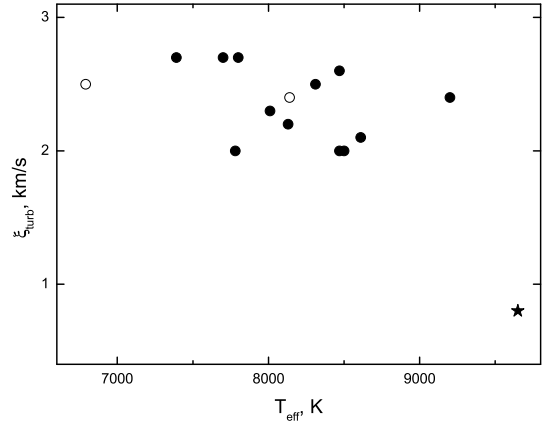,width=10.2cm}
\caption{The dependence of the microturbulence from the effective temperature. The symbols are the same 
as in Figure \ref{feccentr}.}
\label{fvturb}
\end{figure}

We compared our measurements of the projected rotational velocities of the program stars to those obtained
by \citet{am95} and by \citet{rgbgz02}. As it is seen from Figure \ref{fvel} our measurements are in good 
agreement with them. The velocities of \citet{am95} seems to be closer to our values
while those of \citet{rgbgz02} are often slightly higher.  

\begin{figure} 
\epsfig{ file=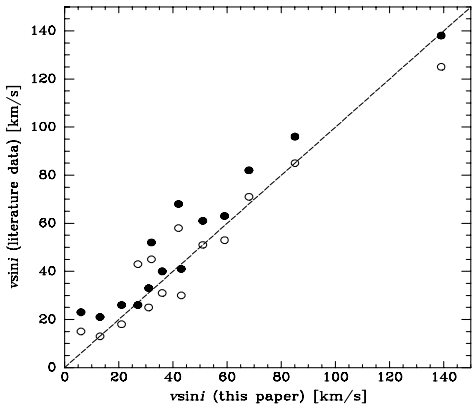,width=9cm}
\caption{The projected rotational velocities obtained by us versus those given by \citet{am95} 
(open circles) and by \citet{rgbgz02} (filled circles).}
\label{fvel}
\end{figure}

\section{Conclusions}
In this third and last paper, we have analysed another six binaries from our sample
and derived their chemical composition, temperatures, gravities, 
projected rotational velocities, masses and ages.
The obtained abundances of certain elements allowed us to conclude that these stars are Am stars.
We suggested that one of them, HD\,155375 could be a new SB2 stars based on 5 spectra obtained.
The tidal interaction in all binary systems studied by us till now was explored. 
There is clear dependence of the Am peculiarity on the projected rotational
velocity for clear defined Am stars. Apart from that there seems to be a correlation of the Am
peculiarity with the eccentricity and may be also with the orbital period and a weak
anticorrelation with the effective temperature.

Assuming that the Am peculiarity is due to the microscopic diffusion
processes in stable atmospheres then this dependence of the Am peculiarity
on the eccentricity must be due to some mechanism which will stabilize the
atmosphere of the star and reduce mixing.

One could speculate that star on an eccentric orbit is subject
to a variable gravitational perturbance or oscillations. These might
cause departures from the single star rotation, drive the star towards the
pseudo-synchronisation, and might reduce the differential rotation and/or
rotationally induced mixing.

This confirms the hypothesis of Budaj (1996,1997) about the existence of a
stabilization mechanism due to the companion of the star and that this
mechanism operates up to the orbital period of about 200 days and that it
is relevant for the Am phenomenon.

\section*{Acknowledgments}
IS and II acknowledge the partial support from Bulgarian NSF under grant D0 02-85. 
JB gratefully acknowledges grant support from the Marie Curie International
Reintegration Grant FP7-200297.

This research was supported partially by the VEGA grants 2/0074/09,
2/0078/10 and 2/0094/11 from the Slovak Academy of Sciences.
This study made use of IRAF Data Reduction and Analysis System and the 
Vienna Atomic Line Data Base (VALD) services.

We wish also to thank the referee P. North for his valuable comments and suggestions
that helped us to improve substantially the paper.

\end{document}